A U-Statistic-based random forest approach for genetic interaction study

Ming Li[1], Ruo-Sin Peng[1], Changshuai Wei[1], Qing Lu[1]

[1]Department of Epidemiology, Michigan State University, East Lansing, MI 48824

**TABLE of CONTENTS**

## 1. ABSTRACT

Variations in complex traits are influenced by multiple genetic variants, environmental risk factors, and their interactions. Though substantial progress has been made in identifying single genetic variants associated with complex traits, detecting the gene-gene and gene-environment interactions remains a great challenge. When a large number of genetic variants and environmental risk factors are involved, searching for interactions is limited to pair-wise interactions due to the exponentially increased feature space and computational intensity. Alternatively, recursive partitioning approaches, such as random forests, have gained popularity in high-dimensional genetic association studies. In this article, we propose a U-Statistic-based random forest approach, referred to as Forest U-Test, for genetic association studies with quantitative traits. Through simulation studies, we showed that the Forest U-Test outperformed exiting methods. The proposed method was also applied to study Cannabis Dependence (CD), using three independent datasets from the Study of Addiction: Genetics and Environment. A significant joint association was detected with an empirical p-value less than 0.001. The finding was also replicated in two independent datasets with p-values of 5.93e-19 and 4.70e-17, respectively.

## 2. INTRODUCTION

The past decade has witnessed an evolutional change of genetic association studies from the research of a limited number of candidate genes to the investigation of entire human genomes. This is made possible by the knowledge of comprehensive high-density maps of single nucleotide polymorphisms (SNPs) and the advancement of genotyping technologies (1-4). These genome-wide association studies (GWASs) have advanced the field of human genetics, allowing us to explore unknown regions for potential risk variants associated with diseases. So far, a large number of GWASs have been conducted and hundreds of novel disease-susceptibility loci have been reported (5, 6). Despite these successes, the current identified loci only explain a small fraction of the diseases' heritability (7). Moreover, the identified genetic variants have a low replication rate in follow-up studies, and have shown a limited contribution to disease prediction (8, 9). Following the initial association scan of the GWASs, it is a natural step for future genetic association studies to explore the potential complex interactions among genetic variants and environmental risk factors (10). Such studies can increase the power to detect unknown genetic variants that are associated with diseases, and provide novel insights into biological pathways underlying the development of diseases (9).

Detecting the gene-gene and gene-environmental interactions has been a longstanding goal of genetic association studies. One strategy is to explore all possible combinations of genetic and environment risk factors and select the best combination predisposing to the risk of disease. Ritchie *et al.* proposed a Multifactor Dimensionality Reduction (MDR) method, which is based on such a strategy (11). For each combination of multiple SNPs, it partitions all possible multi-SNP genotypes into two risk groups, and calculates their classification error for the disease outcome. The combination of SNPs with the lowest classification error is then selected and assessed for joint association with the disease. The limitations of the MDR method are that it requires balanced data and does not allow for covariate adjustments. To address these limitations, Lou *et al.* extended it to a generalized MDR (GMDR) framework (12). Similar as the MDR, the GMDR searches all genetic variants for the 'best' multi-SNP combination. However, it uses score statistics instead of a case control ratio, when partitioning the genotypes into two risk groups. Both the MDR and GMDR methods are non-parametric and model free, and have been successfully applied for detecting the gene-gene and gen-environment interactions associated with complex diseases, such as autism and nicotine dependence (13, 14). However, they both search exhaustively for combinations of SNPs, which are computationally impractical for high-dimensional data (15).

Alternatively, recursive partitioning approaches— such as classification trees and random forests—have been widely adopted for the identification of the gene-gene and gene-environment interactions (16-18). A classification tree is constructed by sequentially adding into the tree model the best predictor among a large number of features. Based on this idea, Li *et al.* proposed a Forward U-Test that examines the combined effect of genetic variants for quantitative traits, with the consideration of gene-gene interactions (19). It first uses a U-Statistic-based forward algorithm to select potential disease-susceptibility loci and then assesses the joint association of the selected loci using U-Statistics. It has been reported that the classification accuracy of a single classification tree can be improved substantially by averaging an ensemble of trees, referred to as random forests (20). A random forest is comprised of multiple decision trees, each of which is grown by using bootstrap samples of the same size. For each decision tree, the approach sequentially selects the best predictor among a subset of randomly selected features, and adds it into the tree model. The random-forest-based methods have been applied successfully to detect gene-gene interactions underlying various complex diseases, such as asthma and age-related macular degeneration (21, 22). In addition to its improved performance, random-forest-based methods have the advantage of detecting high-order interactions on high-dimensional data (23, 24).

So far, random forests have mostly been used in genetic association studies with binary outcomes. In addition, a few linkage analyses have used random forests for the analysis of quantitative traits (25, 26). However, the application of random forests for genetic association studies of quantitative traits is still lacking. Furthermore, the current methods commonly assess the significance level of each SNP using the permutation-based importance scores. Such a procedure is computationally intensive and time demanding. In this article, we propose a Forest U-Test approach for the identification of gene-gene and gene-environmental interactions associated with quantitative traits. The Forest U-Test approach can be looked upon as an extension of the previously developed Forward U-Test with an implementation of random forests. We also derive an asymptotic U-Statistic test to assess the joint association of multiple genetic variants with quantitative traits. Through simulations and a real data application, we compare the proposed Forest U-Test with the existing methods, such as Forward U-Test and GMDR.

## 3. METHODS

Suppose we have a study population of $N$ subjects, each genotyped with $K$ $Y_i$. $X = (X_{i1}, X_{i2} \ldots \ldots X_{iK})$ denote the quantitative trait and observed genotypes for the $i^{th}$ subject, $i = 1, 2, \ldots \ldots, N$. Each SNP may have three possible values, i.e., $X_{ij} \in \{AA, Aa, aa\}$, $j = 1, 2, \ldots \ldots, K$. In the following sub-sections, we first give the definition of U-Statistics and then incorporate it into the random forest process to evaluate the joint association of genetic variants with quantitative traits.

### 3.1. U-Statistics

We have previously introduced a U-Statistic to measure the joint association of multiple SNPs with quantitative traits with the consideration of possible gene-gene interactions (19). Following the same notation, we assume $k$ ($k \leq K$) SNPs are associated with the traits. $L$ multi -SNP genotypes can be formed by these $k$ SNPs, denoted as $G_1, G_2, \ldots \ldots, G_L$. Let $S_l = \{i, X_i = G_l\}$ be the group of subjects carrying multi-SNP genotype $G_l$, and $m_l = |S_l|$ be the number of subjects in $S_l$. We define the between-group U-statistic for group $S_l$ and group $S_{l'}$ as:

$$U_{l,l'} = \sum_{i,j} \varphi(Y_i, Y_j) \quad ; \quad i \in S_l, j \in S_{l'} ;$$

where $\varphi(Y_i, Y_j) = Y_i - Y_j$. 　　　Equation (1)

To measure the overall trait difference among a total of $L$ multi-SNP genotype groups, we further define the global U-statistic as:

$$U = \frac{\sum_{1 \le l < l' \le L} \omega_{l,l'} U_{l,l'}}{\sum_{1 \le l < l' \le L} \omega_{l,l'}} \times \frac{L(L-1)}{2};$$

where $\omega_{l,l'} = \frac{\sqrt{m_l + m_{l'}}}{m_l m_{l'}}$.          Equation (2)

Here, the weight parameter $\omega$ is chosen to account for the number of subjects in each genotype group. When the number of genotype groups is greater than two, we assume that the expected quantitative trait value of the $L$ multi-SNP genotypes decreases with $l$ (i.e., $E(Y_{S_1}) \ge E(Y_{S_2}) \ge \ldots\ldots \ge E(Y_{S_L})$. Practically, we determine the sequence of genotypes groups by ordering their average trait values (i.e., $\overline{Y}_{S_1} \ge \overline{Y}_{S_2} \ge \ldots\ldots \ge \overline{Y}_{S_L}$).

### 3.2. U-Statistic-based decision tree

For common complex traits, the disease susceptibility loci are commonly unknown. In order to determine the $k$ disease-susceptibility loci and the corresponding multi-SNP genotypes, a single decision tree is built using the proposed U-Statistics. We start with a root node comprised of all subjects in the study. In the first step, the root node is split into two offspring nodes comprised of subjects carrying two different single-SNP genotypes. Each SNP $j$ can split the root node in three possible ways, noted as $\{g_1^j = AA, g_2^j = Aa \mid aa\}$, $\{g_1^j = Aa, g_2^j = AA \mid aa\}$ and $\{g_1^j = aa, g_2^j = AA \mid Aa\}$. For each splitting strategy, a U-Statistic can be calculated for two offspring nodes comprised of subjects with genotypes $\{G_1^{(1)} = g_1^j, G_2^{(1)} = g_2^j\}$, where $G_l^{(s)}$ denotes the $l^{th}$ multi-SNP genotype at step $s$. The splitting strategy with the largest U-Statistic is selected. In the second step, a second SNP $j'$, paired with SNP $j$, may form four two-SNP genotypes, denoted by $\{G_1^{(2)} = G_1^{(1)} \wedge g_1^{j'},$ $G_2^{(2)} = G_1^{(1)} \wedge g_2^{j'}, G_3^{(2)} = G_2^{(1)} \wedge g_1^{j'},$ $G_4^{(2)} = G_2^{(1)} \wedge g_2^{j'}\}$. A global U-Statistic is calculated for these four offspring nodes using Equation (2), and the one with the largest global U-Statistic is chosen. The offspring nodes are further split in a binary fashion, until a given number of depth $d$ is reached or the offspring nodes cannot be further split. The tree depth $d$ is defined as the total number of times to split offspring (root) nodes. While splitting each offspring node, the SNP and the splitting strategy are chosen to maximize the corresponding U-Statistics. By doing so, a decision tree can be grown.

### 3.3. U-Statistic-based random forest

The performance of the U-statistic-based decision tree can be further improved by constructing a random forest, built on an ensemble of $T$ decision trees. Each decision tree is grown using a bootstrap sample of the study population with the same size, $N$. The selected and non-selected subjects are referred to as in-bag and out-of-bag samples respectively. Each time an offspring (root) node is split, a subset of $p$ SNPs is randomly sampled from the totality of $K$ SNPs. From these $p$ random features, we select one SNP to split the node according to the same SNPs selection and splitting rule described above. The splitting process is continued until a tree depth $d$ is reached, and a decision tree is built. Each decision tree $t$ will result in a series of multi-SNP genotypes, $G_{t,1}, G_{t,2}, \ldots\ldots, G_{t,L_t}$. Let $S_{t,l}$ be the set of subjects carrying $G_{t,l}$ and $m_{t,l} = |S_{t,l}|$. For every subject $i$ with multi-SNP genotype $G_{t,l}$, a predicted trait value can be calculated by averaging the trait values across all subjects in $S_{t,l}$:

$$\widehat{Y}_{t,i} = \sum_{r \in S_{t,l}} Y_r / m_{t,l};$$

$\forall i \in S_{t,l}$;          Equation (3).

In addition, while considering an ensemble of $T$ decision trees, an in-bag trait value for each subject $i$ can be calculated by averaging the predicted trait values across all the decision trees where subject $i$ is an in-bag sample:

$$\overline{Y}_{i,ib} = \frac{\sum_{t=1}^{T} \widehat{Y}_{t,i} \times I(i \in \{in\ bag\ of\ tree\ t\})}{\sum_{t=1}^{T} I(i \in \{in\ bag\ of\ tree\ t\})};$$

Equation (4).

where $I(\cdot)$ is an indicator function. Similarly, an out-of-bag trait value can be calculated for each subject $i$ as:

$$\overline{Y}_{i,oob} = \frac{\sum_{t=1}^{T} \widehat{Y}_{t,i} \times I(i \in \{out\ of\ bag\ of\ tree\ t\})}{\sum_{t=1}^{T} I(i \in \{out\ of\ bag\ of\ tree\ t\})};$$

Equation (5).

### 3.4. Significance level

Hypothesis testing can then be conducted to evaluate the joint association of $K$ genetic variants with the quantitative traits, considering possible interactions. After constructing the random forest described above, we treat each subject as a separated genotype group. Assuming they are sorted by their in-bag trait values, we can calculate a global U-Statistic by using their out-of-bag trait values (Equation (2)). Specifically, the global U-Statistic has the following form:

$$
\begin{aligned}
U = & (\bar{Y}_{1,oob} - \bar{Y}_{2,oob}) + (\bar{Y}_{1,oob} - \bar{Y}_{3,oob}) + \ldots + (\bar{Y}_{1,oob} - \bar{Y}_{N,oob}) \\
& + (\bar{Y}_{2,oob} - \bar{Y}_{3,oob}) + \ldots + (\bar{Y}_{2,oob} - \bar{Y}_{N,oob}) \\
& + \ldots + (\bar{Y}_{N-1,oob} - \bar{Y}_{N,oob}) \\
= & \sum_{1 \le i < j \le N} (\bar{Y}_{i,oob} - \bar{Y}_{j,oob})
\end{aligned}
$$

$;$     Equation (6);

$$\text{where } \bar{Y}_{1,ib} \ge \bar{Y}_{2,ib} \ge \bar{Y}_{3,ib} \ldots \ge \bar{Y}_{N,ib}$$

This U-Statistic can be used to test the joint association of the SNPs with the traits. A null distribution of U-Statistics can be obtained by permuting the trait and applying the same random forest procedure to calculate the U-Statistics. Based on the null distribution, an empirical p-value can be obtained. The empirical p-value accounts for the inflated type I error due to model selection and ordering of subjects.

We also derive an asymptotic test for replicating the initial association in an independent study. We refer to the data from the original study as training data and the data from the follow-up study as testing data. We denote multi-SNP genotypes of each decision tree $t$ by $G_{t,1}$, $G_{t,2}$, ......, $G_{t,L_t}$. Let $S_{t,l}^{Train}$ and $S_{t,l}^{Test}$ be the set of subjects with genotype $G_{t,l}$ in training and testing data respectively, and denote $Y_r^{Train}$ as the $r^{th}$ subject in training data and $Y_i^{Test}$ be the $i^{th}$ subject in the testing data; where $1 \le r \le N_{Train}$ and $1 \le i \le N_{Test}$. For each subject $i$ in $S_{t,l}^{Test}$, we first obtain its corresponding training predicted trait value in the training data:

$$\widehat{Y}_{t,i}^{Train} = \sum_{r \in S_{t,l}^{Train}} Y_r^{Train} / m_{t,l}^{Train} ;$$

$\forall i \in S_{t,l}^{Test}$ ;     Equation (7).

Meanwhile, a testing predicted trait value can be calculated by averaging the observed traits in testing data across all subjects in $S_{t,l}^{Test}$ :

$$\bar{Y}_{t,i}^{Test} = \sum_{r \in S_{t,l}^{Test}} Y_r^{Test} / m_{t,l}^{Test} ;$$

$\forall i \in S_{t,l}^{Test}$

We further average trait values for each subject $i$ over $T$ decision trees,

$$\bar{Y}_i^{Train} = \sum_{t=1}^{T} \widehat{Y}_{t,i}^{Train} / T ;$$

Equation (8).

$$\bar{Y}_i^{Test} = \sum_{t=1}^{T} \widehat{Y}_{t,i}^{Test} / T ;$$

Equation (9).

Again, we treat each subject in the testing set as a separated genotype group. Assuming they are sorted by their training average trait values, we can calculate a global U-Statistic by using the testing average trait value (Equation (6)). Since the sequence of subjects is pre-determined by the training set, we expect the global U-Statistic follows a normal distribution asymptotically with a zero mean under the null distribution. Assuming all subjects in the testing set are independent and have the same variance of $\sigma^2$, the variance of the global U-Statistic can be calculated as (See Appendix for detail):

$$Var(U) = \sigma^2 \sum_{i=1}^{N} (N+1-2i)^2 ;$$

Equation (10).

It should also be noted that, although the above method is illustrated with genotype data, it applies to environmental risk factors with categorical levels as well. For continuous environmental risk factors, we need to first categorize the continuous variables before applying the proposed method.

## 4. RESULTS

### 4.1. Simulation I

In the first simulation, we compared the performance of the proposed method with that of two existing methods, the Forward U-Test and GMDR. The comparison was conducted under various underlying diseases models with different levels of disease complexity. We also incorporated two types of genetic interactions into the disease models, a multiplicative-effect model and a threshold-effect model (Table 1) [27]. We started with a simple disease model comprised of a two-locus multiplicative effect interaction. The second disease model included a two-locus multiplicative effect interaction and two independent SNPs with additive effects. The third disease model included a two-locus multiplicative effect interaction, a two-locus threshold effect interaction and two independent SNPs. The fourth disease model included a two-locus multiplicative effect interaction, a two-locus threshold effect interaction and four independent SNPs. The SNP genotypes were simulated under the assumption of Hardy-Weinberg Equilibrium (HWE). For all SNPs, the minor alleles corresponded to higher traits and the minor allele frequencies were set at 0.3. For each disease model, a number of non-disease related SNPs were also introduced to bring the total number of SNPs to ten. The minor allele frequencies of these noise SNPs were sampled from a uniform distribution ranging from 0.1 to 0.9. For each underlying disease model, we first simulated a reference population with one million subjects. An expected trait $\widehat{y}_i$ was calculated for each subject according to its multi-SNP genotypes at causal loci. The observed quantitative trait for each subject was simulated as: $y_i = \widehat{y}_i + \varepsilon_i$, where $\varepsilon_i \sim N(0,1)$. For each simulation replicate, we randomly selected 1000 subjects from the reference population, and then analyzed the data by using the Forest U-Test, as well as the Forward U-Test and GMDR. While constructing a random forest for the Forest U-Test, we fixed the number of trees at $T = 500$, the number of random features at $p = 3$ and tree depth at $d = 5$. 'Testing Balance Accuracy' was used as testing statistic for GMDR. For each underlying disease model, the simulation was repeated 1000 times. 1000 permutations were generated to form the empirical null distribution. The association was significant if the testing statistic exceeded the 95[th] percentile of its corresponding permuted distribution. The power was then calculated as the probability of detecting the overall association across 1000 replicates. To evaluate type I error, we simulated another set of quantitative traits from a standard normal distribution, which assumes the traits are independent from individuals' genotypes. The same procedure was applied to calculate corresponding type I errors.

**Table 1.** Average trait values for two-locus interaction models

| Multiplicative Effect | | | Threshold Effect | | | |
|---|---|---|---|---|---|---|
| | bb | Bb | BB | | bb | Bb | BB |
| aa | $\alpha$ | $\alpha(1+\mu_{21})$ | $\alpha(1+\mu_{21})(1+\mu_{22})$ | aa | $\alpha$ | $\alpha$ | $\alpha$ |
| Aa | $\alpha(1+\mu_{11})$ | $\alpha(1+\mu_{11})(1+\mu_{21})$ | $\alpha(1+\mu_{11})(1+\mu_{22})$ | Aa | $\alpha$ | $\alpha(1+\mu)$ | $\alpha(1+\mu)$ |
| AA | $\alpha(1+\mu_{11})(1+\mu_{12})$ | $\alpha(1+\mu_{12})(1+\mu_{21})$ | $\alpha(1+\mu_{12})(1+\mu_{22})$ | AA | $\alpha$ | $\alpha(1+\mu)$ | $\alpha(1+\mu)$ |

The simulation results were summarized in Table 2. From the results, we observed power improvement for all methods as the number of causal loci increased. For the simplest disease model with only two causal SNPs, the power of Forward U-Test is close to that of Forest U-Test, and is higher than the power of the GMDR. With the increase of model complexity, the Forest U-Test had the most significant power increase, and outperformed the other two approaches. In all scenarios, the type I errors were properly controlled for all methods. In the simulation study, we fixed the minor alleles at 0.3 for all causal SNPs.

We expect power increase for all methods if the minor allele frequencies were increased.

### 4.2. Simulation II

In the second simulation, we evaluated the performance of Forest U-Test with respect to two pre-determined parameters, the number of random features and the tree depth $d$. The simulation was conducted by using the fourth disease model from simulation I. We also increased the number of irrelevant SNPs to make sure the total number of SNPs was 25. We varied the random feature $p$ from 3 to 12, and the values of $d$ from 2 to 5. In all replicates, the number of trees was set at $T = 500$. For each combination of $p$ and $d$, the simulation was repeated for 1000 times. Following the same procedure as Simulation I, we estimated the power and type I error.

The simulation results were summarized in Table 3. We reported on the power, type I error and average time cost to construct a random forest with 500 trees. The computational time was based on the time of running $R$ programs on a high-performance computer with a dual-core 1.6GHz processor and 4 GB memory. We found the computational time increased with the increase of either $p$ or $d$. The results also showed that the power of Forest U-Test was improved as $d$ increases. In our simulation, the highest power was attained when $p = 8$ and $d = 5$. While $p$ was fixed at 8, limited power improvement was gained when $d$ increased from 4 to 5. With respect to the number of random features, the power of Forest U-Test was first improved as $p$ increased, but was reduced when $p$ reached a larger value ($p = 12$). In all scenarios, the type I error was properly controlled.

### 4.3. Simulation III

In the simulation, we compared the proposed approach with the Random Forest (RF) approach developed by Breiman L. *et al.* (20). RF does not provide a testing statistic for association testing. Instead, it focuses on variable selection and ranks all SNPs by importance score. Therefore, we compared the performance of two approaches by their importance ranking of all SNPs. The performance of two approaches was evaluated based on the fourth disease model in simulation I. For both RF and Forest U-Test, 500 trees were constructed. The parameters were fixed at $p = 3$ and $d = 5$. The RF analysis was conducted by using package 'randomForest_4.6-2' in R (28). While applying RF, the importance scores, measured by the mean decrease of MSE (i.e., the default measurement in RF), is used to rank the SNPs. While applying Forest U-Test, the SNPs were ranked by the selection times to construct decision trees. The result was summarized in Figure 1.

**Table 2.** Comparison between Forest U-Test and Forward U-Test/GMDR under various disease models

| Disease Models | | Forest U-Test | Forward U-Test | GMDR |
|---|---|---|---|---|
| Two-locus Multiplicative [1] | Power | 0.338 | 0.294 | 0.049 |
| | Type I Err. | 0.053 | 0.065 | 0.039 |
| Two-locus Multiplicative + Two Independent Loci [2] | Power | 0.696 | 0.472 | 0.147 |
| | Type I Err. | 0.037 | 0.042 | 0.047 |
| Two-locus Multiplicative + Two-locus Threshold [3] + Two Independent Loci | Power | 0.772 | 0.505 | 0.184 |
| | Type I Err. | 0.058 | 0.055 | 0.051 |
| Two-locus Multiplicative + Two-locus Threshold + Four Independent Loci | Power | 0.886 | 0.622 | 0.192 |
| | Type I Err. | 0.043 | 0.043 | 0.058 |

[1]For all SNP blocks with multiplicative effects, the parameters of average trait values were set as $\mu_{11} = \mu_{21} = 1.1$, $\mu_{12} = \mu_{22} = 1.2$, [2]For all SNP blocks with threshold effects, the parameters of average trait values were set as $\mu = 1.2$, [3]For all independent loci, the average trait values for genotype aa, Aa, AA were set as 1, 1.1 and 1.2 respectively

**Table 3.** Performance of the Forest U-Test with various parameters

| d / p | | 2 | 3 | 4 | 5 |
|---|---|---|---|---|---|
| 3 | Power | 0.553 | 0.592 | 0.644 | 0.693 |
| | Type I Err. | 0.033 | 0.044 | 0.049 | 0.046 |
| | Time | 0.92 min | 1.81 min | 3.03 min | 4.75 min |
| 6 | Power | 0.600 | 0.647 | 0.695 | 0.742 |
| | Type I Err. | 0.033 | 0.038 | 0.037 | 0.035 |
| | Time | 1.24 min | 2.36 min | 3.83 min | 5.84 min |
| 8 | Power | 0.666 | 0.703 | 0.751 | 0.759 |
| | Type I Err. | 0.037 | 0.037 | 0.037 | 0.036 |
| | Time | 1.96 min | 3.75 min | 6.05 min | 9.19 min |
| 12 | Power | 0.540 | 0.590 | 0.642 | 0.685 |
| | Type I Err. | 0.026 | 0.031 | 0.033 | 0.036 |
| | Time | 3.38 min | 6.36 min | 10.08 min | 14.8 min |

In Figure 1, we found that both methods selected causal SNPs (SNP 1 – SNP 8) with relatively higher importance. We also found that RF ranked the causal SNPs with threshold effect as less important than the causal SNPs with multiplicative effect or independent effect. On the other hand, the ranking of the causal SNPs by Forest U-Test was consistent across different modes of inheritance. This may be partially due to the splitting strategy of RF. During the tree constructing, RF can choose different SNPs to split nodes at the same level, which might be a disadvantage for capturing interaction models, in particular the threshold effect model. Suppose the first SNP can split the root node into two offspring nodes as a 'risk' allele node and a 'non-risk' allele node. The second SNP is less likely to be selected by RF to further split the 'non-risk' allele node, which lead to the low chance of capturing threshold effect interactions. On the other hand, Forest U-Test uses the same SNP to split nodes at the same level, and hence, increase the power to capture interactions.

### 4.4. Application to cannabis dependence

We applied the proposed method to study Cannabis Dependence by using the Study of Addiction: Genetics and Environment (SAGE) GWAS dataset (29). As part of the Gene Environment Association (GENEVA) consortium (30), SAGE was designed by selecting unrelated participants from three independent studies: the Family Study of Cocaine Dependence (FSCD), the Collaborative Study on the Genetics of Alcoholism (COGA), and the Collaborative Genetic Study of Nicotine Dependence (COGEND). In our analysis, the trait of interest is Cannabis Dependence, measured by the number of marijuana symptoms endorsed (mj_sx_tot). The trait has eight numerical values, ranging from 0 to 7. The distribution of traits is given in Figure 2. We also collected 25 SNPs that had been reported in the previous literatures as having potential association with Cannabis Dependence. The genotypes of 13 SNPs were available in SAGE GWAS dataset and the genotypes of the remaining 12 SNPs were imputed by software package IMPUTE2 (31, 32). The CEU and YRI populations from the HapMap phase III and 1000 Genome project were used as the reference panels to impute the 12 SNPs for white and black subjects (33, 34). In addition to the 25 SNPs, gender was also included in the analysis as a covariate.

We used FSCD as an initial association dataset, and COGA and COGEND as replicate datasets. While applying the Forest U-Test, we set the parameters as $p = 8$, $d = 10$, and $T = 500$. The results were listed in Table 4. In the initial dataset of FSCD, a strong joint association of 25 SNPs and gender with Cannabis Dependence has been detected. Permutation test was conducted to account for an inflated type I error due to model selection and subject ordering. The empirical p-value of the association was <0.001. Evaluation of the joint association in COGA (p-value=5.93e-19) and COGEND (p-value=4.70e-17) showed the association remained highly significant. We also listed top SNPs which were essential for constructing the random forest. The trained random forest was comprised of 500 decision trees, and gender was selected 476 times as the most important covariate. The three top SNPs were rs2501432 (C/T), rs324420 (A/C) and rs1431318 (C/T), which located in genes *CNR2*, *FAAH* and *ANKFN1*, respectively. They were selected 382, 366 and 364 times, respectively.

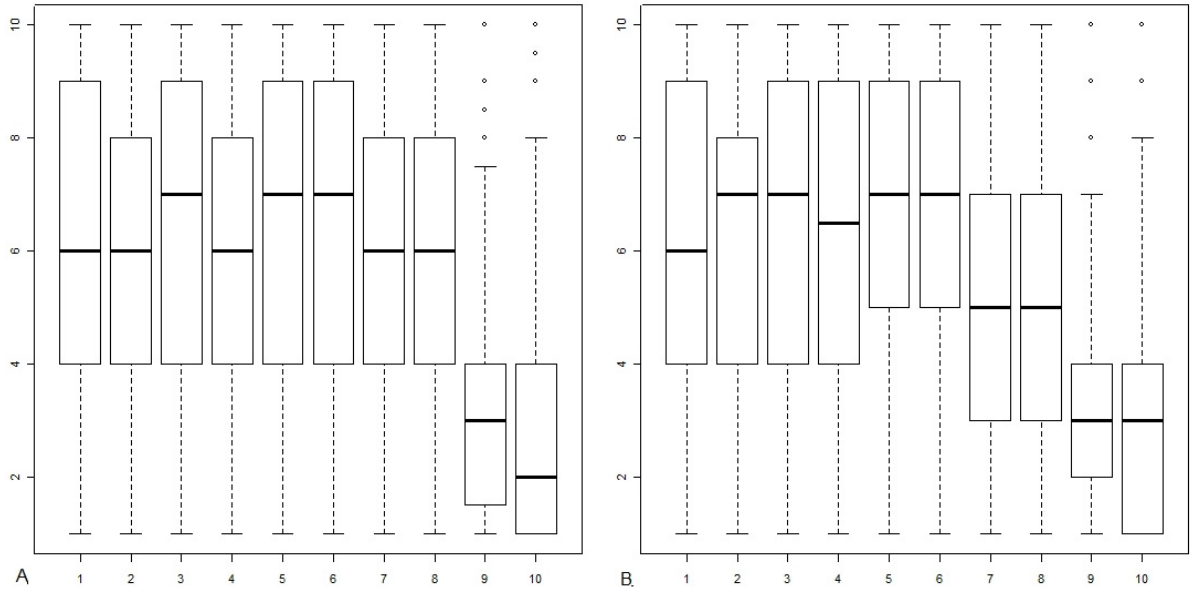

**Figure 1.** Comparison between Forest U-Test and RF by the box plots of importance ranks. A. Forest U-Test ；     B. RF In both plots, 10 SNPs were considered: SNP 1 - SNP 4 : four causal SNP with independent effect；  SNP 5 - SNP 6:  two causal SNPs with multiplicative joint action; SNP 7 – SNP 8: two causal SNPs with threshold joint action; SNP 9 – SNP 10: two non-causal SNPs.

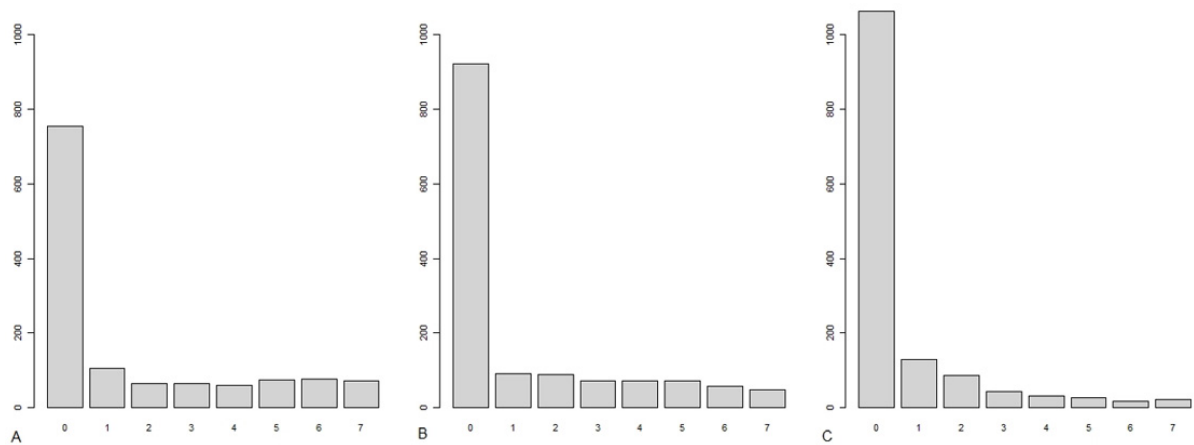

**Figure 2.** Distribution of the trait in SAGE datasets. A. Trait distribution in FSCD. B. Trait distribution in COGA. C. Trait distribution in COGEND.

We also applied the Forward U-Test and GMDR to the same datasets. Similar to the above analysis, FSCD was used as initial dataset for model selection, and COGA and COGEND were used for replication. The results were listed in Table 5 and Table 6. The analysis result of FSCD showed that both the Forward U-Test and GMDR selected the most parsimonious models with gender only. Therefore, the analyses of COGA and COGEND only examined the association between the trait and gender. Though the association remained significant in all studies, neither method detected any genetic effects.

To compare the Forest U-Test with the conventional RF approach, we also applied RF to the same datasets. When RF was applied to FSCD, gender was also ranked as the most important covariate. The three top SNPs were rs2501432,

rs1019238 and rs324420. These three SNPs were ranked as 1st, 6th and 3rd important SNPs by Forest U-Test. While the findings of both methods were highly consistent, it was not straightforward for RF to replicate its initial finding in COGA and COGEND. When RF was applied to COGA and COGA, gender remained to be the most important covariates. However, the top SNPs changed across studies (Table 7).

## 5. DISCUSSION

Evidence has shown that multiple genes are interacting in biological pathways to influence the development of diseases (9, 35). It is also common for genetic effects to be modified by environmental risk factors (36). Ignoring the complex interactions between genes and environmental factors will likely reduce the power of detecting novel risk factors underlying complex traits (37). Though there is an increasing awareness that the gene-gene and gene-environment interactions are crucial for understanding the etiology of complex diseases, detecting the complex interactions remains a major challenge in genetic association studies (15, 38). Recently, random forest approaches have been adopted to detect the association of multiple risk factors while allowing for high order interactions. However, most of the current studies have applied random forests to binary disease outcomes, and few studies have considered their applications for quantitative traits. In this article, we propose a U-Statistic-based random forest approach for genetic association studies with quantitative trait. The proposed method was found to have a greater power than two existing methods, the Forward U-Test and GMDR. The simulation results showed that the Forest U-Test had the largest advantage over the existing methods when the underlying disease model is highly complex. This improvement can be explained by the following reasons: 1) By constructing an ensemble of decision trees based on random features and bootstrap samples, the method not only considers the risk factors with large effects, but also incorporate those with only small or moderate effects. Though many risk factors may only play a limited role in the disease development, they can collectively contribute to a significant portion of the variation of traits. On the other hand, the Forward U-Test and GMDR only search the risk factors for the best combination and may overlook those with small or moderate effects; 2) Compared to a single decision tree, the random forests provide a more robust performance, making the result replicable in the follow-up studies; 3) By averaging the predicted trait values of multiple decision trees, Forest U-Test allows for a large number of risk groups (i.e. every subject may form a risk group). On the contrary, the Forward U-Test only allows for a limited number of risk groups formed by a few selected risk factors, while GMDR always assumes two risk groups for each combination of risk factors. Such assumptions may be questionable in real disease scenarios.

**Table 4.** Association result of Forest U-Test in FSCD and its replication in COGA and COGEND

| Study | Top Covariates | Allele | Chro | Gene | p-values |
|---|---|---|---|---|---|
| FSCD | Gender rs2501432 rs324420 rs1431318 | C/T | 1 | CNR2 | <0.001 |
| | | A/C | 1 | FAAH | |
| | | C/T | 17 | ANKFN1 | |
| COGA | Replicate Study for model trained by FSCD | | | | 5.93e-19 |
| COGEND | Replicate Study for model trained by FSCD | | | | 4.70e-17 |

**Table 5.** Association result of Forward U-Test in FSCD and its replication in COGA and COGEND

| Study | Covariates | p-values |
|---|---|---|
| FSCD | Gender | <0.001 |
| COGA | Replicate Study with Gender | 1.56e-14 |
| COGEND | Replicate Study with Gender | 3.08e-17 |

**Table 6.** Association result of GMDR in FSCD and its replication in COGA and COGEND

| Study | | Model | Training  Bal. Acc | Testing  Bal. Acc | Sign Test (p) | CV |
|---|---|---|---|---|---|---|
| FSCD | 1 | Gender | 0.6138 | 0.6162 | 9 (0.0107) | 10/10 |
| | 2 | Gender  rs1049353 | 0.6156 | 0.5894 | 9 (0.0107) | 3/10 |
| | 3 | Gender  rs1045642  rs2501432 | 0.6302 | 0.5665 | 7 (0.1719) | 4/10 |
| COGA | | Gender (replicate study) | 0.6392 | 0.6397 | 10 (0.001) | 10/10 |
| COGEND | | Gender    (replicate study) | 0.6570 | 0.6553 | 10 (0.001) | 10/10 |

**Table 7.** Analysis result of RF in FSCD, COGA and COGEND

| Study | Top Covariates | | | |
|---|---|---|---|---|
| FSCD | Gender | rs2501432 | rs1019238 | rs324420 |
| COGA | Gender | rs4680 | rs1019238 | rs324420 |
| COGEND | Gender | rs1431318 | rs1049353 | rs2070744 |

We also note that the Forest U-Test evaluates the joint association of multiple risk factors without directly selecting the most parsimonious combination of risk factors. Therefore, compared to the Forward U-Test and GMDR, the results of Forest U-Test are less easy for interpretation, which is a common limitation for most random-forest-based methods. Nevertheless, the asymptotic result of Forest U-Test can easily be used to evaluate the association in independent follow-up studies, which is an advantage over the conventional random-forest-based methods.

Cannabis Dependence is a disorder that may involve complex interactions among multiple risk factors. In our analysis, the three top SNPs come from three genes CNR2, FAAH, and ANKFN1 respectively. CNR2, also known as Cannabinoid receptor type 2, belongs to the cannabinoid receptor family. The encoded protein functions as a receptor for cannabinoids, which are the principal psychoactive ingredients of marijuana (39). This gene has been verified to occur in the central nervous system, and is expressed in the brain (40, 41). SNP rs2501432 is a non-synonymous mutation that locates in CNR2. Experimental evidence has demonstrated that this mutation can change the function of CNR2 protein (42). Previous studies have also reported the association of the SNP with substance use disorders (43, 44). SNP rs324420, also known as Pro129Thr, is another non-synonymous mutation that locates in exon 3 of the fatty acid amide hydrolase gene (FAAH), which has shown associations with many substance use disorders (45-47). It was estimated that the minor allele homozygote leaded to a reduced risk of 0.25 for the development of Cannabis Dependence (48). The third SNP, rs1431318, lies in gene *ANKFN1*, which was previously identified as being involved in substance use disorders (49, 50). Interestingly, this SNP was reported as the most significant SNP (p-value <1e-7) by a recent GWAS for Cannabis Dependence (51). Whereas it is biologically plausible that all these genes may play an important role in developing Cannabis Dependence, our method does not provide any inference for the underlying biological mechanism. The identified association may result from either additive or interactive effects between the genes. Future study would be necessary to further replicate the association and investigate potential joint actions among these genes.

## 6. ACKNOWLEDGEMENT

This work is supported by start-up funds from Michigan State University. The authors want to thank four referees for helpful comments that improved this manuscript.

**Appendix:** Derivation of the variation of U-Statistics in Equation (7). Note that when each subject is treated as a separated group, the global U-Statistic in Equation (2) is equivalent to

$$U = \sum_{i=2}^{N}(Y_1 - Y_i) + \sum_{i=3}^{N}(Y_2 - Y_i) + \ldots\ldots + \sum_{i=N}^{N}(Y_{N-2} - Y_i) + (Y_{N-1} - Y_N)$$

$$= (N-1)Y_1 + (N-2)Y_2 - Y_2 + (N-3)Y_3 - 2Y_3 + \ldots\ldots + (1-N)Y_N$$

$$= \sum_{i=1}^{N}[(N-i)Y_i - (i-1)Y_i]$$

$$= \sum_{i=1}^{N}(N+1-2i)Y_i$$

Assuming $Y_i$ are mutually independent and $Var(Y_i) = \sigma^2$, we have

$$Var(U) = \sigma^2 \sum_{i=1}^{N}(N+1-2i)^2$$

**Send correspondence to:** Qing Lu, Department of Epidemiology, Michigan State University, East Lansing, MI 48824 USA, Tel: 517-353-8623, Fax: 517-432-1130, E-mail: qlu@epi.msu.edu